\documentclass[aps,prl,reprint,10.5pt]{revtex4-1}

\usepackage{amsmath,amsthm,amsfonts,amssymb,bm}
\usepackage{mdframed,graphicx,array,MnSymbol}
\usepackage[usenames,dvipsnames]{xcolor}
\definecolor{OceanBlue}{rgb}{0,0.35,0.7} 
\usepackage{hyperref}
\hypersetup{colorlinks=true,allcolors=OceanBlue}

\theoremstyle{definition}

\theoremstyle{remark}
\newmdenv[
  topline=false,
  bottomline=false,
  rightline=false,
  skipabove=\topsep,
  skipbelow=\topsep,
  leftmargin=-5pt,
  rightmargin=-10pt,
  innertopmargin=0pt,
  innerbottommargin=0pt
]{leftrule} 

\newcommand{\pars}[1]{\left(#1\right)} 
\newcommand{\bracs}[1]{\left[#1\right]} 

\newcommand{\mb}{\mathbf} 
\newcommand {\cl}{\mathcal} 
\newcommand{\vectsym}{\boldsymbol}  
\DeclareMathOperator{\tr}{tr\,}            
\DeclareMathOperator{\Tr}{Tr\,}            

\newcommand{\bphi}{\vectsym{\phi}} 

\newcommand {\ket}[1] {\left|{#1}\right\rangle}
\newcommand{\varket}[1] {\left|{#1}\right\rrangle}
\newcommand{\varbra}[1] {\left\llangle #1\right|}
\newcommand{\varbraket}[2] {\left\llangle #1| #2 \right\rrangle}

\newcommand {\bra}[1] {\langle{#1}|}

\newcommand{\abs}[1]{\left | #1 \right |}   

\begin{document}
\title{Quantum-limited loss sensing: Multiparameter estimation and Bures distance between loss channels}
\author{Ranjith Nair}
\email{nairanjith@gmail.com}
\affiliation{Department of Electrical and Computer Engineering, \\ National University of Singapore, 4 Engineering Drive 3, 117583 Singapore\\
}

\date{\today}
\begin{abstract} 
The problem of estimating multiple loss parameters of an optical system using the most general ancilla-assisted  parallel strategy is solved under energy constraints. An upper bound on the quantum Fisher information matrix is derived assuming that the environment modes involved in the loss interaction can be accessed.  Any pure-state probe that is number-diagonal in the modes interacting with the loss elements is shown to exactly achieve this upper bound even if the environment modes are inaccessible, as is usually the case in practice. We explain this surprising phenomenon, and show that measuring the Schmidt bases of the probe is a  parameter-independent  optimal measurement. Our results imply that multiple copies of two-mode squeezed vacuum probes with an arbitrarily small nonzero degree of squeezing, or probes prepared using single-photon states and linear optics can achieve quantum-optimal performance in conjunction with on-off detection. We also calculate explicitly the energy-constrained Bures distance between any two product loss channels. Our results are relevant to standoff image sensing,  biological imaging, absorption spectroscopy, and  photodetector calibration.
\end{abstract}
\maketitle

\begin{figure}[h]
\centering\includegraphics[trim=70mm 32mm 40mm 20mm, clip=true,width=\columnwidth]{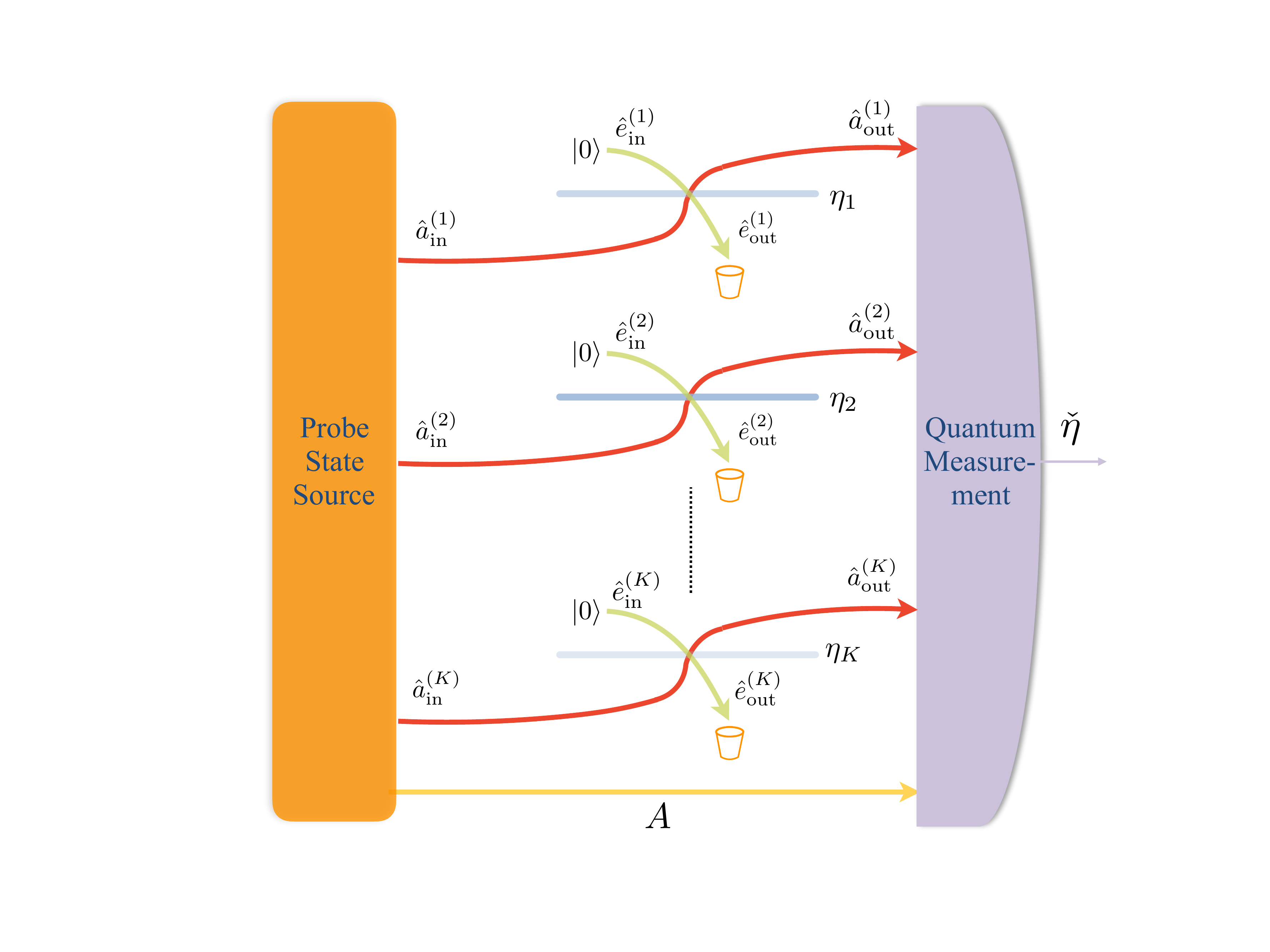}
\caption{
General ancilla-assisted  parallel  strategy for loss sensing: A joint probe state of the signal (red) and ancilla (yellow) modes $A$ is prepared. Each signal mode queries one of $K$ beam splitters with unknown transmittances $\vectsym{\eta} = (\eta_1,\ldots \eta_K)$. The output signal modes and ancilla modes are jointly measured to yield an estimate $\check{\vectsym{\eta}}$ of $\vectsym{\eta}$. The input environment modes (green) entering the beam splitters are in the vacuum state,  the output environment modes are inaccessible, and multiple signal modes (not shown) may query each beam splitter.
} \label{fig:estimationsetup}
\end{figure}

Quantum metrology investigates the fundamental limits imposed by quantum mechanics on the precision of measurements under resource constraints. It encompasses the measurement of such physical quantities  as displacement, time, force, acceleration, temperature, and electric and magnetic fields in diverse physical systems including atoms, ions, and spins  \cite{GLM11,*TA14,*DRC17,BAB+18}. 
In optical systems, measuring interferometric phase shifts  has been the paradigmatic example for which quantum enhancement has been most studied \cite{Cav81,D-DJK15}. 

Beyond unitary dynamics, the most ubiquitous phenomenon present in optical systems is loss, the precise measurement of which is a fundamental issue in science and technology. 
A general loss sensing scenario is depicted schematically in Fig.~\ref{fig:estimationsetup}. $K$  loss elements modeled as beam splitters  are probed using a multimode probe of which the ``signal''  modes (shown in red) are directly modulated by the loss elements while the ``ancilla''  modes (shown in yellow) are held losslessly. The exact nature of the modes and loss elements need not be specified for our analysis, which applies to diverse scenarios. Thus the $K$ loss elements  may be actual pixels in an amplitude mask  in an image sensing scenario \cite{BGB10},  or may represent  absorption coefficients of a sample at $K$ different frequencies in an absorption spectroscopy setup \cite{WCN+17}, or a photodetector whose quantum efficiency is being calibrated \cite{ABD+11,*QJL+16,*MLS+17,*S-CWJ+17,*MS-CW+17}. The $K$ probes may also represent temporal modes probing the transmittance of a living cell undergoing a cellular process \cite{TB16}. More abstractly, many natural imaging problems  can be mapped to equivalent transmittance estimation or discrimination problems  \cite{TNL16,*NT16prl,*LP16,Hel73,*LKN+18arxiv}. At optical frequencies, we may assume that the environment modes entering the ``unused'' input ports of the beam splitters are  in the vacuum state. The output environment modes are typically inaccessible for measurement, so only the signal and ancilla modes are measured using an optimal quantum measurement in order to estimate the transmissivity values. In order to make the problem well-defined, we constrain the  energy \footnote{Throughout this paper, I refer to the average  photon number of a state of a given set of quasimonochromatic modes simply as its  ``energy''.} allocated to the signal modes. Apart from accounting for resources, it is often necessary  to limit the photon flux through an optical element, e.g., to avoid damage or alteration of processes in live tissue \cite{Col14}, to calibrate sensitive single-photon detectors \cite{ABD+11}, or for covertness. 

In this Letter, we solve the problem of quantum-optimal estimation of $K$ real-valued transmissivities $\{\sqrt{\eta}_k\}_{k=1}^K$ using the general ancilla-assisted entangled parallel strategy of Fig.~\ref{fig:estimationsetup} assuming that the energy allocated to the signal modes probing each of the $K$ beam splitters is specified as $\{N_k\}_{k=1}^K$. We first obtain an upper bound on the quantum Fisher information matrix for the problem, and then show that a very large class of probe states achieves this bound. We find the optimal quantum measurement and exhibit readily prepared probes for which on-off detection is an optimal measurement. Finally, we address the problem of discriminating between two given loss channels by deriving the probe state of given signal energy that minimizes the fidelity at the channel output.

\emph{Problem Formulation and Estimation-Theory Review:}
The action of the $k$-th beam splitter on the $m$-th signal mode annihilation operator $\hat{a}^{(m)}$ and the $m$-th environment mode annihilation operator $\hat{e}^{(m)}$ takes the form
\begin{equation}
\begin{aligned} \label{HPevolution}
\hat{a}^{(m)}_{\rm out} &= \sqrt{\eta_k}\;\hat{a}^{(m)}_{\rm in}+ \sqrt{1-\eta_k}\;\hat{e}^{(m)}_{\rm in}, \\
\hat{e}^{(m)}_{\rm out} &= \sqrt{1- \eta_k}\;\hat{a}^{(m)}_{\rm in} - \sqrt{\eta_k}\;\hat{e}^{(m)}_{\rm in}
\end{aligned}
\end{equation}
in the Heisenberg picture, where $1 \leq m \leq M \equiv \sum_{k=1}^K M_k$,  with $M_k$ the total number of signal modes probing the $k$-th loss element. In the Schr\"odinger picture, this evolution is generated by the unitary operator
\begin{align} \label{SPevolution}
\hat{U}^{(m)}(\phi_k) = \exp\bracs{-i\phi_k\pars{\hat{a}^{(m)\dag}\hat{e}^{(m)} + \hat{a}^{(m)}\hat{e}^{(m) \dag}}},
\end{align}
where  the ``angle'' parameter $\phi_k \in [0,\pi/2]$ satisfies $\cos \phi_k= \sqrt{\eta_k}$. For an initial vacuum environment state, the evolution defines a quantum channel  $\cl{L}_{\eta_k}$ on the signal mode that maps an input state $\rho$ to the output state $\cl{L}_{\eta_k}(\rho)$ with Wigner characteristic function
\begin{align} \label{lossmap}
\chi_{\rm out}(\xi) = \chi_{\rm in}(\sqrt{\eta}_k \,\xi)\, e^{-(1-\eta_k)\abs{\xi}^2/2}\mbox{ for } \xi \in \mathbb{C},
\end{align}
where $\chi_{\rm in}(\xi) = \Tr \rho \,e^{\xi\hat{a}_{\rm in}^{(m)\dag} - \xi^* \hat{a}_{\rm in}^{(m)}}$ is the characteristic function of $\rho$.

Without loss of generality, we assume that the  signal $(S)$ and ancilla $(A)$ modes are in a joint pure state $\ket{\psi}$ (viz., the \emph{probe}) satisfying the energy constraints $\bra{\psi}\hat{N}_k \ket{\psi} = N_k$ for $k=1,\ldots, K$, where $\hat{N}_k = \sum_{M_k\,{\rm modes}} \hat{a}^{(m)\dag} \hat{a}^{(m)}$ is the total photon number operator of the signal modes probing the $k$-th loss element. Including the environment ($E$) modes (initially in the multimode vacuum state $\ket{\mb 0}_E$) and the ancilla modes, the output state of the total system $ASE$ is  given by $\sigma_{\bphi} = \ket{\Psi_{\bphi}}\bra{\Psi_{\bphi}}_{ASE}$, where
\begin{align} \label{ASEstate}
\ket{\Psi_{\bphi}}_{ASE} = \hat{I}_A \otimes  \pars{\otimes_{k=1}^K \otimes_{m=1}^{M_k} \hat{U}^{(m)}(\phi_k)} \ket{\psi}_{AS}\ket{\mb{0}}_E,
\end{align}
$\hat{I}_A$ is the identity operator on the ancilla system, and $\bphi = (\phi_1,\ldots, \phi_K)$.  Since the output environment modes are actually inaccessible, the measured output state  is $\rho_{\bphi} = \Tr_E \sigma_{\bphi}$. 

The state family $\{\rho_{\bphi}\}$ gives rise to the corresponding multi-parameter quantum Cram\'er-Rao bound (QCRB) \cite{Hel76,*Hol11,GLM11,Par09}. Briefly, for each parameter $\phi_i$, there exists a Hermitian operator $\hat{L}_i$ (that depends on $\bphi$ in general) called the symmetric logarithmic derivative (SLD)  satisfying $\partial_i \rho_{\bphi} \equiv \partial \rho_{\bphi}/ \partial \phi_i = \pars{\rho_{\bphi} \hat{L}_i + \hat{L}_i \rho_{\bphi}}/2$. The quantum Fisher information matrix (QFIM) $\cl{K}$  is the $K \times K$ matrix with $ij$-th matrix element  given by
$\cl{K}_{ij} =  \Tr \rho_{\bphi} \pars{\hat{L}_i\hat{L}_j + \hat{L}_j\hat{L}_i}/2.$ 
 Consider any measurement applied to the output modes resulting in an \emph{estimate} vector $\check{\bphi} = \left (\check{\phi}_1,\ldots, \check{\phi}_K \right)$ for $\bphi$. The error covariance matrix $\Sigma$ of the estimate has the matrix elements $\Sigma_{ij} = \mathbb{E} \bracs{\pars{\check{\phi}_i - \phi_i}\pars{\check{\phi}_j - \phi_j}}$, where $\mathbb{E}$ denotes statistical expectation over the measurement results. For an unbiased estimate, i.e., if $\mathbb{E}\bracs{\check{\phi}_i} = \phi_i$ for all $\bphi$ and $i$,  the QCRB is the matrix inequality 
$\Sigma \geq \cl{K}^{-1}$. For any $K \times K$ positive semidefinite cost matrix $G$, the QCRB implies that the scalar cost  $\tr G \Sigma \geq \tr G \cl{K}^{-1}$ for any unbiased estimator $\check{\bphi}$ \footnote{For example, taking $G=I$ yields a lower bound on  the sum of the mean squared errors of the parameters.}. The SLD operators in terms of the transmittance parametrization $\vectsym{\eta} = (\eta_1, \ldots, \eta_K)$ are related to those of the angle parametrization via  $\hat{L}_k^{(\vectsym{\eta})} = \frac{\partial \phi_k}{\partial \eta_k}\, \hat{L}_k$ and result in a different QFIM and QCRB. We will indicate which parametrization is being used by a subscript or superscript where necessary.

The estimation of a single loss parameter has been studied before \cite{SM06,VF07,MP07,ADD+09,MI10,MI11,CDB+14} (see \cite{BAB+18} for a review), but not in the generality considered here. Thus, ref.~\cite{SM06} focused on measurement optimization, \cite{VF07}  on  specific probes and measurements, \cite{MP07} studied single-mode Gaussian-state probes, while \cite{ADD+09} considered optimizing the state of a single-signal-mode probe. Refs.~\cite{MI10} and \cite{MI11} studied ancilla-assisted schemes using Gaussian probes, while \cite{CDB+14} studied joint estimation of loss and phase using a single signal-ancilla mode pair.  Thus, none of these works addressed the  general multimode ancilla-assisted parallel strategy for multi-parameter loss estimation. 

\emph{Upper bound on the QFIM} --
We first obtain an upper bound (in the matrix-inequality sense) on the QFIM for estimating $\bphi$, extending the approach of Monras and Paris \cite{MP07} for the single-parameter case. Suppose that the output environment modes $\left\{\hat{e}_{\rm out}^{(m)}\right\}_{m=1}^M$ are accessible. From Eq.~\eqref{ASEstate}, the purified output state $\sigma_{\bphi} = e^{-i\sum_k \phi_k \hat{H}_k}\,\sigma\,e^{i\sum_k \phi_k \hat{H}_k}$ for $\sigma = \ket{\psi}\bra{\psi}_{AS}\otimes\ket{\mb{0}}\bra{\mb{0}}_E$ and $\hat{H}_k =  \hat{I}_A\otimes\pars{\sum_{m=1}^{M_k}\hat{a}^{(m)\dag}\,\hat{e}^{(m)} + \hat{a}^{(m)}\,\hat{e}^{(m)\dag}}$. Since $\sigma_{\bphi}$ is pure, differentiating $\sigma_{\bphi}^2 =\sigma_{\bphi}$ implies that an SLD operator $\hat{\ell}_k$ for $\phi_k$ is $\hat{\ell}_k = 2\partial_k \sigma_{\bphi} = 2i\bracs{\sigma_{\bphi}, \hat{H}_k}$. A direct calculation of the $ij$-th matrix element of the QFIM $\widetilde{\cl{K}}_{ij}= \Tr \sigma_{\bphi} \pars{\hat{\ell}_i\hat{\ell}_j + \hat{\ell}_j\hat{\ell}_i}/2$ (the tilde denotes that this matrix is calculated assuming access to the environment modes) gives
$
\widetilde{\cl{K}}_{ij} = 4 \,\bra{\psi}\pars{ \sum_{m=1}^{M_k} \hat{a}^{(m)\dag}\,\hat{a}^{(m)} }\ket{\psi}\, \delta_{ij} = 4N_i\, \delta_{ij},
$
where $N_i$ is the total energy of $\ket{\psi}$ in the signal modes probing the $i$-th  beam splitter. The monotonicity of the QFIM \cite{[{}][{, Theorem 10.3.}]Pet08qits} under partial trace over $E$  implies that the true QFIM matrix $\cl{K}_{\bphi}$ satisfies 
\begin{align} \label{genMPlimit}
\cl{K}_{\bphi} \leq  \widetilde{\cl{K}}_{\bphi} = 4 \; {\rm{diag}}(N_1,\ldots, N_K).
\end{align}
Note that this bound is valid for any probe state with the given signal energy distribution and is independent of the values of $\{M_k\}_{k=1}^K$. We refer to \eqref{genMPlimit} as the \emph{generalized Monras-Paris (MP) limit}.

\emph{The performance of NDS probes}: We now exhibit probes saturating the limit \eqref{genMPlimit}. Consider first the  case of a single beam splitter probed by a pure joint signal-ancilla state $\ket{\psi}$ with $M$ signal modes. It is easily seen that any such  $\ket{\psi}$ can be written as 
\begin{align} \label{probe}
\ket{\psi} = \sum_{\mb{n} \geq \mb{0}} \sqrt{p}_{\mb{n}}\ket{\chi_{\mb n}}_A \ket{\mb n}_S,
\end{align}
where $\ket{\mb n}_S = \ket{n_1}_{S_1}\ket{n_2}_{S_2}\cdots \ket{n_M}_{S_M}$ is an $M$-mode number state of $S$, $\{\ket{\chi_{\mb n}}_A\}$ are normalized (not necessarily orthogonal) states of $A$, and $p_{\mb n}$ is the probability distribution of $\mb{n}$. 
The energy constraint takes the form
\begin{align} \label{ec}
&\sum_{n=0}^{\infty} n\,p_n = N;\;\;{\rm for}\;\; p_n = \sum_{\mb{n}\, :\, n_1 +\ldots + n_M = n} p_{\mb n}
\end{align}
the probability mass function of the total photon number in the signal modes.  For any transmittance value  $\eta (\phi)$, we can write the output state \eqref{ASEstate} explicitly as
\begin{align} \label{}
\ket{\Psi_\phi}_{ASE} &=  \sum_{\mb{l} \geq \mb{0}}  \varket{\psi_{\phi;\mb{l}}}_{AS} \ket{\mb{l}}_E,
\end{align}
where 
\begin{align} \label{lcondstates}
\varket{\psi_{\phi;\mb{l}}}_{AS} = \sum_{\mb{n} \geq\mb{l}}  \sqrt{p_\mb{n} B_\eta(\mb{n},\mb{l})} \ket{\chi_{\mb{n}}}_A\ket{\mb{n}-\mb{l}}_S
\end{align} 
are unnormalized states of $AS$, $B_\eta(\mb{n},\mb{l}) = \Pi_{m=1}^M {n_m \choose l_m} \eta^{n_m - l_m} (1-\eta)^{l_m}$ is a product of binomial probabilities, and $\mb{n} \geq \mb{l}$ is to be understood component-wise. Since $\varbraket{\psi_{\phi;\mb{l}}}{\psi_{\phi';\mb{l}}}\geq 0$, the fidelity between the purified output states corresponding to a pair of values $\eta (\phi)$ and $\eta' (\phi')$ equals
\begin{align} \label{ASEfidelity}
F\pars{\sigma_\phi,\sigma_{\phi'}} \equiv \Tr\sqrt{\sqrt{\sigma_\phi} \sigma_{\phi'} \sqrt{\sigma_\phi}}= \sum_{\mb{l}\geq\mb{0}} \varbraket{\psi_{\phi;\mb{l}}}{\psi_{\phi';\mb{l}}}.
\end{align} 
The reduced state of the $AS$ system is given by $\rho_\phi =  \Tr_E \sigma_\phi = \sum_{\mb{l}} \varket{\psi_{\phi;\mb{l}}} \varbra{\psi_{\phi;\mb{l}}}$. Probes $\ket{\psi}$ for which the $\{\ket{\chi_{\mb n}}_A\}$ are orthonormal are called \emph{Number Diagonal Signal (NDS) states} \cite{Nai11,NY11} since the reduced state on $S$ is then number-diagonal. For such probes, we have $\varbraket{\psi_{\phi;\mb{l}}}{\psi_{\phi';\mb{l'}}} = \varbraket{\psi_{\phi;\mb{l}}}{\psi_{\phi';\mb{l}}} \delta_{\mb{l},\mb{l'}}$ from \eqref{lcondstates}, and the output fidelity evaluates to (see also \cite{Nai11} for the latter equality)
\begin{align} \label{NDSfidelity}
F\pars{\rho_\phi,\rho_{\phi'}} = \sum_{\mb{l}\geq\mb{0}} \varbraket{\psi_{\phi;\mb{l}}}{\psi_{\phi';\mb{l}}} =  \sum_{n=0}^{\infty} p_n \mu^n,
\end{align}
where $\mu = \sqrt{\eta\, \eta'} + \sqrt{(1-\eta)(1-\eta')} = \cos\,(\phi'-\phi) \in [0,1]$. Significantly, the middle expression equals the fidelity \eqref{ASEfidelity} between the output states on $ASE$. The QFI can now be calculated as
\begin{align} \label{QFIfromfidelity}
\cl{K}_{\phi} = -4 \,{\partial^2 F\pars{\rho_\phi,\rho_{\phi'}}}/{\partial \phi'^2}\big\vert_{\phi'=\phi} = 4N,
\end{align}
where the first equality is a general relation between fidelity and QFI \cite{Hay06,*BC94}, and the latter follows from \eqref{NDSfidelity}.  We have thus shown that for NDS probes, the QFI with or without environmental access is equal to the MP limit for all values of $\phi$, $N$, and $M$.

For $M=1$ and integer $N$, taking $p_N=1$ in \eqref{probe} recovers the number-state optimality result of \cite{ADD+09}.  However, for non-integer $N$ and in particular for $N < 1$, the unentangled states proposed in \cite{ADD+09} are suboptimal if ancilla entanglement is allowed.  The two-mode squeezed vacuum (TMSV) state is an NDS probe and its  QFI was computed by different techniques \cite{MI11}, although its optimality was not pointed out. Our result  implies that these are just two examples among an infinity of optimal probes.

For the multiparameter case $\bphi = (\phi_1, \ldots, \phi_K)$ with the given energy budget $\left\{N_k\right\}$, consider the product probe state $ \otimes_{k=1}^K \rho^{(k)}$ for $\rho^{(k)} = \ket{\psi^{(k)}}\bra{\psi^{(k)}}$ with $\ket{\psi^{(k)}}$ any NDS probe of signal energy $N_k$. For the resulting output state family $\left\{\rho_{\bphi} = \otimes_{k=1}^K \rho_{\phi_k}^{(k)}\right\}$, it is readily seen that the $k$-th SLD operator $\hat{\Lambda}_k = \hat{I} \otimes \cdots \otimes \hat{L}_k \otimes \cdots \otimes \hat{I}$ for $\hat{L}_k$ satisfying $\partial_k \rho_{\phi_k}^{(k)} = \pars{\hat{L}_k\,\rho_{\phi_k}^{(k)} + \rho_{\phi_k}^{(k)}\, \hat{L}_k}/2$. The SLDs are commuting and the $ij$-th element of the QFIM is:
\begin{align} \label{QFIMelement}
\pars{\cl{K}_{\bphi}}_{ij} = \Tr \rho_{\bphi} \,\hat{\Lambda}_i\,\hat{\Lambda}_j = 4N_i\,\delta_{ij},
\end{align}
where we have used $\Tr \rho_{\phi_k}^{(k)} \hat{L}_k = 0$ for all $k$ and the single-parameter result $\Tr \rho_{\phi_k}^{(k)} \hat{L}_k^2 = 4 N_k$. Thus, the product probe $\otimes_{k=1}^K \rho^{(k)}$ achieves the generalized MP limit \eqref{genMPlimit}. Since the SLDs commute and the QFIM is diagonal, there is no obstacle to the simultaneous achievement of the QCR bounds for the parameters \cite{Hel76,RJD-D16}.

In the transmittance parametrization, the optimal QFIM is given by $\cl{K}_{\vectsym{\eta}}= {\rm diag} \pars{\frac{N_1}{\eta_1(1-\eta_1)}, \cdots,\frac{N_K}{\eta_K(1-\eta_K)}}.$
In comparison, the QFIM for a product coherent-state input with the given energies is  $\cl{K}_{\vectsym{\eta}}^{\rm CS} =  {\rm diag} \pars{{N_1}/{\eta_1}, \cdots,{N_K}/{\eta_K}}$ so that a large advantage is available if $\eta_k \simeq 1$.

\emph{Optimal measurement and practical probes} -- For any state family $\{\rho_\phi\}$,  measuring the basis corresponding to the SLD  achieves the QFI \cite{BC94}, but this basis may be parameter-dependent and thus of limited use \cite{B-NG00}. For a single parameter and an NDS probe of the form \eqref{probe}, consider jointly measuring at the output its Schmidt bases, i.e., the basis $\left\{ \ket{\chi_{\mb{n}}}_A\right\}$ on $A$ and the number basis $\left\{ \ket{\mb{n}}_S \right\}$ on $S$. Such a measurement yields a pair $(\mb{N},\mb{Q})$, where $\mb{N}$ denotes the index of the measured $\ket{\chi_{\mb{n}}}_A$ and $\mb{Q}=\mb{N} - \mb{L}$ is the measured photon number in $S$. The classical Fisher information on $\eta$ of this measurement is
\begin{align} \label{FIschmidtbasis}
\cl{J}_\eta = \sum_{\mb{n}} p_{\mb{n}} \sum_{\mb{l} \leq \mb{n}} B_\eta(\mb{n},\mb{l}) \bracs{\frac{\partial \ln B_\eta(\mb{n},\mb{l})}{\partial \eta}}^2 = \frac{N}{\eta(1-\eta)},
\end{align}
so that the QFI is attained for any $\eta$.

In many sensing scenarios, the values of $\{M_k\}_{k=1}^K$ are not fixed beforehand but can be varied (e.g., by using multiple temporal or spatial modes). This flexibility allows the design of practical probes and measurements. Thus, the NDS probe
\begin{equation}
\begin{aligned} \label{opt1photonstate}
&\ket{\psi_{N}}= \ket{1}_{S_1}\otimes \cdots \otimes \ket{1}_{S_{\lfloor N \rfloor}} \\
&\otimes \pars{\sqrt{1 - \{N\}}\ket{1}_A  \ket{0}_{S_{\lceil N \rceil}}+ \sqrt{\{N\}}\ket{0}_A\ket{1}_{S_{\lceil N \rceil}}}
\end{aligned}
\end{equation} with $M = \lceil N \rceil$ signal modes is optimal for estimating a single loss parameter and can be prepared using single-photon sources and linear optics. Here $\{N\} = N - \lfloor N \rfloor$ is the fractional part of $N$. The optimal measurement described above reduces to on-off detection in each of the $S$ and $A$ modes, making the overall scheme realizable with single-photon technologies \cite{EFM+11}. Similarly, using $M$ copies of a TMSV state (which is also NDS) with each component having signal energy $N/M$ attains the MP limit with an arbitrarily small level of squeezing per mode in the limit of large $M$. In the limit of small squeezing, on-off detection in every mode again becomes a quantum-optimal measurement.

\begin{figure}[tbp]
\centering\includegraphics[trim=90mm 100mm 100mm 75mm, clip=true,width=\columnwidth]{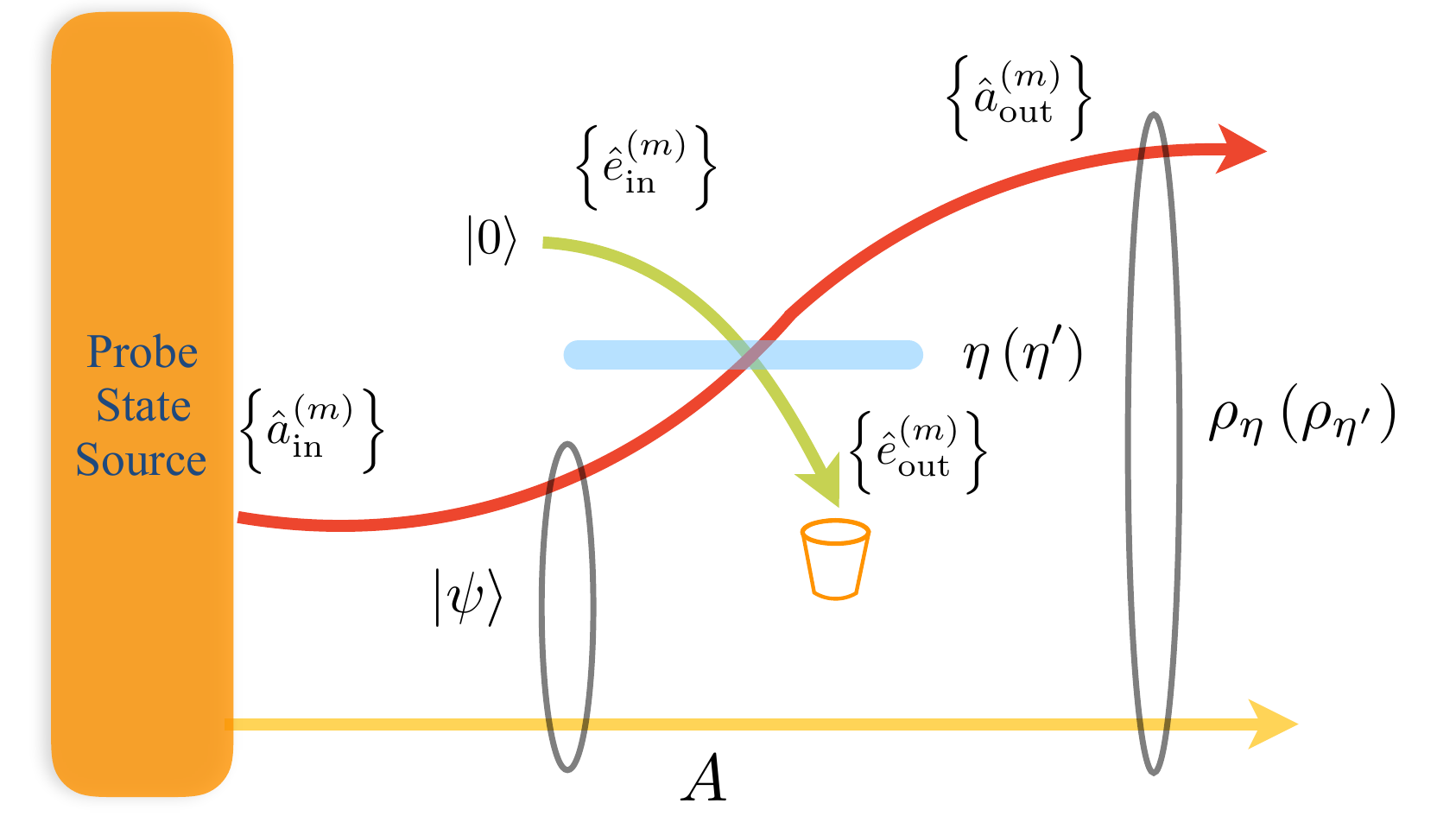}
\caption{
Setup pertaining to the energy-constrained Bures distance between loss channels $\cl{L}_{\eta}^{\otimes M}$ and $\cl{L}_{\eta'}^{\otimes M}$: Each of $M$ signal modes (red) of a probe state $\ket{\psi}$ of  Eq.~\eqref{probe} entangled with an arbitrary ancilla system $A$ (yellow) with total signal energy $N$  queries a beam splitter with transmissivity either $\sqrt{\eta}$ or $\sqrt{\eta'}$. We wish to minimize the fidelity between the respective output states $\rho_\eta$ and $\rho_{\eta'}$ of the signal-ancilla system. 
} \label{fig:ecbdsetup}
\end{figure}

\emph{Energy-constrained Bures distance between loss channels} --  Consider the ancilla-assisted channel discrimination problem  shown in Fig.~\ref{fig:ecbdsetup} in which a probe of a given number $M$ of signal modes and total energy $N$ entangled with an  ancilla $A$ queries a  beam splitter in order to determine which of  two possible values of transmissivity (angle) $\sqrt{\eta}\; (\phi)$ and $\sqrt{\eta'} \;(\phi')$ it possesses.   This problem arises naturally in the quantum reading of a digital optical memory \cite{Pir11} where the loss channels represent bit values. Several measures of general channel distinguishability under an energy constraint have been proposed recently, e.g.,  two versions of the energy-constrained diamond distance \cite{PLO+17,Shi18,Win17arxiv,PL17},  the \emph{energy-constrained Bures  (ECB) distance} \cite{Shi16arxiv}, and general energy-constrained channel divergences \cite{SWA+18}. The ECB distance between any bosonic channels $\cl{M}$ and $\cl{N}$ on the signal modes $S$ is given by the expression \footnote{My formulation \eqref{ecbdistance} differs slightly from that in \cite{Shi16arxiv} in using an equality energy constraint, and is normalized to lie between 0 and 1. As I show, the two definitions give the same ECB distance up to normalization.}:
\begin{equation}
\begin{aligned} \label{ecbdistance}
&B_N(\cl{M},\cl{N}) := \\
& \sup_{\ket{\psi} : \bra{\psi} \hat{I}_A\otimes  \hat{N}_S \ket{\psi} =N} \sqrt{1 - F\pars{{\rm id}\otimes\cl{M}(\ket{\psi}\bra{\psi}),{\rm id}\otimes\cl{N}(\ket{\psi}\bra{\psi})}},
\end{aligned}
\end{equation}
where $F$ is the fidelity, $A$ is an arbitrary ancilla system, id is the identity channel on $A$, $\ket{\psi}$ is a probe  of the form \eqref{probe}, $\hat{N}_S$ is the total photon number operator on $S$, and the optimization is over all pure states of $AS$ with signal energy $N$. 

We now evaluate $B_N\pars{\cl{L}_{\eta}^{\otimes M},\cl{L}_{\eta'}^{\otimes M}}$ between loss channels of the form of Eq.~\eqref{lossmap}.  We first note  that among probes with given $\{p_\mb{n}\}$, the fidelity between the outputs of the channels is lower-bounded by the NDS value $F =\sum_{n=0}^{\infty} p_n \mu^n$  \cite{Nai11}.
Thus,  we need to minimize  $\sum_n p_n \mu^n$ under the energy constraint \eqref{ec}. For an arbitrary $\{ p_n\}$ satisfying the energy constraint, let ${A}_{\downarrow} = \sum_{n \leq \lfloor N \rfloor} p_n$, and $A_{\uparrow} = 1 - {A}_{\downarrow}$. For ${N}_{\downarrow} = A_{\downarrow}^{-1} \sum_{n \leq \lfloor N \rfloor} n\,p_n \leq \lfloor N \rfloor$ and 
${N}_{\uparrow} = A_{\uparrow}^{-1}\sum_{n \geq \lceil N \rceil} n\,p_n \geq \lceil N \rceil$, we have   $A_\downarrow\,N_\downarrow + A_\uparrow\,N_\uparrow = N$. Since the function $x \mapsto \mu^x$ is convex, we have
$F =  \sum_n p_n \mu^n \geq {A}_{\downarrow} \,\mu^{N_{\downarrow}} + {A}_{\uparrow} \,\mu^{N_{\uparrow}}$.
Convexity of this function also implies that the chord joining $({N}_{\downarrow}, \mu^{N_{\downarrow}})$ and $({N}_{\uparrow}, \mu^{N_\uparrow})$
lies  above that joining $(\lfloor N \rfloor, \mu^{\lfloor N \rfloor})$ and $(\lceil N \rceil , \mu^{\lceil N \rceil})$ in the interval $\lfloor N \rfloor \leq x \leq \lceil N \rceil$.  Since the energy constraint  can  be satisfied by taking ${N}_\downarrow = \lfloor N \rfloor, {N}_\uparrow = \lceil N \rceil, p_{\lfloor N \rfloor} = 1 -\{N\}$, and $p_{\lceil N \rceil} = \{N\}$ \footnote{The strict convexity of $x \mapsto \mu^x$ implies that this is the only signal photon number distribution achieving minimum fidelity. Thus, Gaussian probes are  suboptimal.}, the energy-constrained minimum fidelity equals:
\begin{align} \label{ecminf}
F^{\rm min}_N = \pars{1- \{N\}}\,\mu^{\lfloor N \rfloor} +  \{N\}\,\mu^{\lceil N \rceil}.
\end{align}
The ECB distance $B_N\pars{\cl{L}_{\eta}^{\otimes M},\cl{L}_{\eta'}^{\otimes M}}$ then follows from Eq.~(16). Since the  ECB distance is an increasing function of $N$, it equals (up to normalization) the ECB distance defined using an inequality constraint in \cite{Shi16arxiv}. 

The $M$-signal-mode NDS probe 
\begin{equation}
\begin{aligned}
\ket{\psi_N^{\rm ecb}} &=  \sqrt{1 -\{N\}}\ket{\chi_{\lfloor N \rfloor}}_A\ket{\lfloor N \rfloor}_{S_1} \ket{0}_{S_2}\cdots\ket{0}_{S_M} \\
+& \sqrt{\{N\}}\ket{\chi_{\lceil N \rceil}}_A\ket{\lceil N \rceil}_{S_1}\ket{0}_{S_2}\cdots\ket{0}_{S_M}
\end{aligned}
\end{equation}
is optimal for any orthogonal ancilla states $\ket{\chi_{\lfloor N \rfloor}}_A$ and $\ket{\chi_{\lceil N \rceil}}_A$ since it has the optimal total signal photon number distribution. Note that $B_N\pars{\cl{L}_{\eta}^{\otimes M},\cl{L}_{\eta'}^{\otimes M}}$ is independent of $M$, depends on the loss values through $\phi' - \phi$ alone, and that the above state achieves it regardless of these values.

\emph{Discussion} --
Our NDS-probe optimality results imply the  optimality of the quantum imaging scheme of \cite{BGB10} and of the absolute calibration method \cite{Kly80,*JR86,*HJS99,*WM04,ABD+11,*QJL+16,*MLS+17,*S-CWJ+17,*MS-CW+17} employing TMSV probes. For multimode TMSV probes with small per-mode squeezing, and for the probe \eqref{opt1photonstate}, the optimality of on-off detection obviates the need for photon counting using cryocooled detectors.

The surprising result that NDS probes  attain the exact same performance that access to the output environment modes would give contrasts with the case of estimating Hamiltonian shift parameters in the presence of noise, for which the performance is strictly worse than the noiseless case even with ancilla entanglement \cite{EdMFD11,*DKG12,*Tsa13}.   NDS probes are known to be optimal in the global Bayesian approach for general lossy image estimation problems \cite{NY11}. Our results call for a more general investigation into their optimality within the local QFI-based approach , and for other bosonic channels. It remains to be seen if sequential adaptive estimation strategies \cite{D-DM14,CMW16,Yua16, TW16arxiv,PL17,PLL18arxiv} can yield still more quantum enhancement. 

 The exact expression for the ECB distance between loss channels contrasts with the available results for unitary channels, which are in the form of quantum-speed-limit bounds \cite{Fre16}. The ECB distance  gives two-sided bounds on the energy-constrained diamond distance \cite{FvdG99} and hence on the error probability of quantum reading \cite{Pir11}. It has also been used to characterize the fidelity of continuous-variable quantum gates \cite{SW18arxiv}. NDS probes are known to optimize general energy-constrained channel divergences between any two phase-covariant bosonic channels  \cite{SWA+18}. It is thus hoped that other energy-constrained channel divergences may be calculated using similar methods and their metrological consequences be explored.
  
\begin{acknowledgments}
I thank C.~Lupo, S.~Pirandola, M.~E.~Shirokov, and M.~M.~ Wilde for useful discussions. This work is supported by the  Singapore Ministry of Education Academic Research Fund Tier 1 Project R-263-000-C06-112.
\end{acknowledgments}
\bibliography{LossEstbib}
\bibliographystyle{apsrev4-1}
\end{document}